\documentclass[9pt]{article}  \usepackage{times}
\usepackage{graphicx}

\usepackage{color}

\usepackage[round,numbers,sort&compress]{natbib} 

\usepackage{amsmath}
\usepackage{amssymb}
\usepackage{stackrel}
\usepackage{chemarrow}
\usepackage{graphicx}
\usepackage{textgreek}

\usepackage{booktabs}
\usepackage{dcolumn}
\usepackage{rotating}
\usepackage{multirow}

\usepackage[british]{babel}
\usepackage{stfloats}

\renewcommand\appendix{\par
  \setcounter{section}{0}
  \setcounter{subsection}{0}
  \setcounter{figure}{0}
  \setcounter{table}{0}
  \renewcommand\thesection{Appendix \Alph{section}}
  \renewcommand\thefigure{\Alph{section}\arabic{figure}}
  \renewcommand\thetable{\Alph{section}\arabic{table}}
}

\begin{document}

\twocolumn[{\LARGE \textbf{The free energy of biomembrane and nerve excitation and the role of anesthetics\\*[0.2cm]}}
{\large Tian Wang, Tea Mu\v{z}i\'{c}, Andrew D. Jackson and Thomas Heimburg$^{\ast}$\\*[0.1cm]
{\small Niels Bohr Institute, University of Copenhagen, Blegdamsvej 17, 2100 Copenhagen \O, Denmark}\\*[-0.1cm]

{\normalsize \textbf{ABSTRACT}\hspace{0.5cm} In the electromechanical theory of nerve stimulation, the nerve impulse consists of a traveling region of solid membrane in a liquid environment. Therefore, the free energy necessary to stimulate a pulse is directly related to the free energy difference necessary to induce a phase transition in the nerve membrane. It is a function of temperature and pressure, and it is sensitively dependent on the presence of anesthetics which lower melting transitions. We investigate the free energy difference of solid and liquid membrane phases under the influence of anesthetics. We calculate stimulus-response curves of electromechanical pulses and compare them to measured stimulus-response profiles in lobster and earthworm axons. We also compare them to stimulus-response experiments on human median nerve and frog sciatic nerve published in the literature. 
\\*[0.3cm] }}
\noindent\footnotesize{\textbf{Keywords:} phase transitions, nerves, stimulus-response profiles solitons, anesthesia, action potential \\*[0.1cm]}
\noindent\footnotesize {$^{\ast}$corresponding author, theimbu@nbi.ku.dk. }\\
\vspace{0.3cm}
]



\normalsize
\section{Introduction}
\label{sec:Introduction}

The nervous impulse is most commonly recognized as a phenomenon related to current and voltage changes \cite{Johnston1995}. 
It is less known that the nerve pulse also has a mechanical \cite{Tasaki1980, Iwasa1980a, Iwasa1980b, GonzalezPerez2016} and a thermal component \cite{Abbott1958, Howarth1968, Howarth1975, Ritchie1985, Tasaki1989}. Based on the latter findings, it has been proposed that the nerve pulse possesses features of a solitary electromechanical pulse \cite{Heimburg2005c, Heimburg2007b, Heimburg2008, Andersen2009, Lautrup2011}. The theory is based on the nonlinear (electro-) mechanical properties caused by the melting transition of biological membranes from an ordered gel to a disordered fluid state \cite{Heimburg2007a, Muzic2018}. In particular, membranes display a larger lateral compressibility in a transition \cite{Heimburg1998}.  In biological membranes, this transition typically occurs in a temperature regime 10-15$^\circ$ below physiological temperature \cite{Heimburg2007a}. Since this phase transition depends on temperature, pressure, voltage and the chemical potential of drugs such as anesthetics, the free energy necessary to induce it depends on these intensive variables. Interestingly, it is well-known that changes in these variables either excite nerves or inhibit them. Determining the free energy required to excite a membrane transition is the topic of this paper.

Gel membranes display a larger volume and area density than fluid membranes. For this reason, a transition in a membrane from fluid to gel can be induced by changes in hydrostatic \cite{Bottner1994, Czeslik1998, Ebel2001} or lateral pressure. The latter has been extensively studied on monolayers \cite{Albrecht1978, McConnell1984, Brockman1994}, where a liquid expanded to liquid condensed phase transition is found with increasing lateral pressure. Close to transitions, the response of the membrane to changes in lateral pressure is non-linear. This gives rise to the emergence of density pulses in membranes which resemble solitons or solitary waves \cite{Heimburg2005c, Heimburg2007b, Lautrup2011}. Such pulses have been found experimentally in monolayers close to transitions \cite{Griesbauer2012a, Griesbauer2012b, Shrivastava2014, Shrivastava2015}, and electromechanical behavior has also been observed in living nerves \cite{Iwasa1980a, Iwasa1980b, Tasaki1980, Tasaki1989, GonzalezPerez2016}. 

The features of these solitary pulses depend largely on the properties of the melting transition in the membranes and in particular on the distance of the transition from physiological temperature. Any variable that influences the position of the transition relative to physiological conditions potentially affects the excitability of the membrane \cite{Heimburg2007c, Graesboll2014}. Besides changes in ambient temperature, this applies most notably to changes in hydrostatic or lateral pressure, pH, calcium and other salt concentrations \cite{Trauble1976, Jahnig1976} and the addition of small drugs such as anesthetics \cite{Kharakoz2001, Heimburg2007c}, which are all known to influence the transition temperature.

The example of general anesthetics is of particular interest. It has long been known that general anesthetics obey the Meyer-Overton correlation \cite{Overton1901, Overton1991, Cantor1997a, Kharakoz2001, Urban2002}. This correlation expresses the fact that the potency of an anesthetic is proportional to its solubility in the lipid membrane of cells. It can be expressed as $P\cdot [ED_{50}]=$const., where $P$ is the partition coefficient of the anesthetic in the membrane, and $[ED_{50}]$ is the free anesthetic concentration for which 50\% of all individuals are anesthetized \cite{Heimburg2007c}. The critical dose of anesthetics in the lipid membrane is thus the same for all general anesthetics. This has important consequences, e.g., the known additivity of the effect of different anesthetics. The Meyer-Overton correlation is an empirical finding that does not in itself represent an explanation. The fact that anesthetics dissolve in membranes does not imply a mechanism. However, it has been shown that the Meyer-Overton correlation acquires a physical meaning if one takes into account the fact that general anesthetics dissolve only in fluid membranes and not in gel membranes \cite{Kharakoz2001, Heimburg2007c}. It has been shown that general and local anesthetics induce a melting point depression in the membrane that is described by \cite{Heimburg2007c, Graesboll2014}
\begin{equation}\label{eq_Intro01}
\Delta T_m=-\frac{RT_m^2}{\Delta H}x_A
\end{equation}
where $T_m$ is the melting temperature, $\Delta H$ is the latent heat of the phase transition, and $x_A$ is the molar fraction of anesthetics in the lipid membrane. At critical dose, $x_A$ and $\Delta T_m$ are always the same ($\approx 2.7$ mol\% and -0.6$^\circ$, respectively  \cite{Heimburg2007c}). 

The lowering of the melting temperature by anesthetics can be compensated by the application of hydrostatic pressure \cite{Johnson1950, Halsey1975, Trudell1975, Ueda1999a, Heimburg2007c}. For tadpoles, the critical pressure at which anesthetized tadpoles recover is about 50 bars. It can be calculated from the known pressure dependence of the melting temperature of lipid membranes \cite{Heimburg2007c}, which suggests that the melting transition is somehow coupled to the function of nerves.

The temperature, pressure, lateral pressure and anesthetic dependence of the chemical potential difference between fluid and gel membrane,  $\mu_{gf}(T, \Delta p, \Delta \Pi, x_A)$, is given by \cite{Heimburg2007c} :

\begin{eqnarray}\label{eq_Intro02}
\Delta \mu_{gf} &\approx& \Delta H\left( \frac{T_m-T}{T_m}+\gamma_V \Delta p \frac{T}{T_m}\right.\nonumber\\
&&\left. +\gamma_A \Delta \Pi \frac{T}{T_m}-\frac{RT}{\Delta H}x_A+...\right) 
\end{eqnarray}
where $\gamma_V=7.8\cdot 10^{-10}$ m$^2$/J (for DPPC) and $\gamma_A=0.89$ m/J are coefficients related to volume and area expansion of the lipids in the transition. The above equation contains a term for each intensive variable. For this reason, the free energy difference also depends on the transmembrane voltage \cite{Heimburg2012, Mosgaard2015a}, even though the exact functional form of the term remains to be explored. 

In the soliton theory for nerve pulse propagation, the pulse consists of a segment of solid phase traveling in the otherwise liquid membrane. The free energy for exciting a pulse is therefore related to the chemical potential difference for the conversion of fluid membrane to gel membrane. 

The free energy difference between a fluid and a gel membrane is given by
\begin{equation}\label{eq_Intro03}
\Delta G_{gf}=f\cdot \Delta\mu_{gf}\;,
\end{equation}
where f is the fraction of fluid lipid.
The free energy necessary to convert a membrane at $T, p, \Pi, x_A$ from fluid to gel is therefore given by
\begin{equation}\label{eq_Intro04}
\Delta G_{ex}=-\Delta G_{gf} \;.
\end{equation}
This is the free energy required to excite a soliton. It depends on the intensive variables, e.g., voltage, lateral pressure, temperature or the chemical potentials of anesthetics. The dependence on anesthetic concentration will be explored below.

\section*{Materials and Methods}
\textbf{Chemicals:}
Lidocaine (98\%), NaCl, KCl, CaCl$_2$, MgCl$_2$, Tris, and glucose were purchased from Sigma-Aldrich. All water used was Milli Q water (18.1 M$\Omega$) prepared by a Direct-Q\textsuperscript{\textregistered} 3 UV water purification system (Merck). \\*[0.1cm]
\noindent\textbf{Porcine spine membranes:}
Porcine spine was acquired from the local butcher. It was kept in a refrigerator and was never frozen. The spinal cord was homogenized using a rotor-stator 125 Watt Lab Homogenizer with 7 mm Probe (Tissue Master, Omni International Inc., Kennesaw, GA) at 33.000 RPM in 30-second intervals for about 30 minutes. The homogenized tissue was diluted with a 150mM NaCl, 10mM Na$_2$HP0$_4$ and 1,8 mM KH$_2$P0$_4$ buffer at pH 7.4. The homogenized sample was filtrated through stainless steel 100 mesh with 140 \textmu m opening size (Ted Pella, Inc., Redding, CA) in order to remove large fibers. Samples were centrifuged at 20800 RCF for 10 min in an Eppendorf desk centrifuge. After centrifugation one finds a pallet of membranes. The supernatant was discarded in order to remove soluble proteins and other soluble components. The removed supernatant was replaced with buffer, vortexed and centrifuged again. This procedure was repeated five times. The pallet was assumed to contain the clean spinal cord membranes, which is evident from the heat capacity profile containing a pronounced lipid melting peak. More details can be found in \cite{Muzic2016}.\\*[0.1cm]
\noindent\textbf{Earthworm preparations:}
Earthworms (Lumbricus terrestris) were obtained from a local supplier. The procedure for extracting the nerve is described in detail in \cite{GonzalezPerez2014}. We used an earthworm saline solution adapted from \cite{Drewes1974} consisting of 75 mM NaCl, 4 mM KCl, 2 mM CaCl$_2$, 1 mM MgCl$_2$, 10 mM Tris, and 23 mM glucose, adjusted to pH 7.4.  \\*[0.1cm]
\noindent\textbf{Lobster preparations:}
Lobsters \textit{Homarus americanus} were acquired from a local supplier. Nerves were extracted following a procedure described in \cite{GonzalezPerez2014}. We used a lobster saline solution adapted from \cite{Evans1976} (462 mM NaCl, 16 mM KCl , 26 mM CaCl$_2$, 8 mM MgCl$_2$, 10 mM TRIS and 11 mM Glucose, adjusted to pH 7.4.\\*[0.1cm]
\noindent\textbf{Nerve recordings:}
Extracellular recordings of the action potential were performed using a Powerlab 26T (ADInstruments, Australia), an integrated data acquisition system with built-in function generator. Details are given in \cite{GonzalezPerez2014}.\\*[0.1cm]
\noindent\textbf{Calorimetry:}
Calorimetric scans were performed on a MicroCal VP-DSC (Northampton, MA) differential scanning calorimeter. Heating rates were kept at 20$^\circ$ C/hour, filtering period was 5 seconds and feedback was set to none. Prior to the experiment, samples were degassed for a few minutes under light vacuum.


\section*{Theory}
\subsection*{Electromechanical solitons}
The fluctuation-dissipation theorem implies that fluctuations in enthalpy are proportional to the heat capacity, while fluctuations in volume and area are proportional to their respective compressibilities \cite{Heimburg1998}. Furthermore, volume, area and enthalpy are proportional functions of temperature \cite{Heimburg1998, Pedersen2010}. This convenient fact allows us to calculate the compressibility of membranes from the excess heat capacity. One finds that the compressibility is a proportional function of the heat capacity. The sound velocity in a membrane is given by $c=1/\sqrt{\kappa_S\cdot \rho^A}$, where $\kappa_S$ is the adiabatic compressibility and $\rho^A$ is the area density of the lipid membrane \cite{Heimburg1998, Halstenberg1998, Schrader2002}. For low frequencies, the isothermal and the adiabatic compressibility are approximately equal \cite{Mosgaard2013a}, and the sound velocity can also be determined from the heat capacity \cite{Heimburg2005c}. 
\begin{figure}[htb!]
	\begin{center}
		\includegraphics[width=8.5cm]{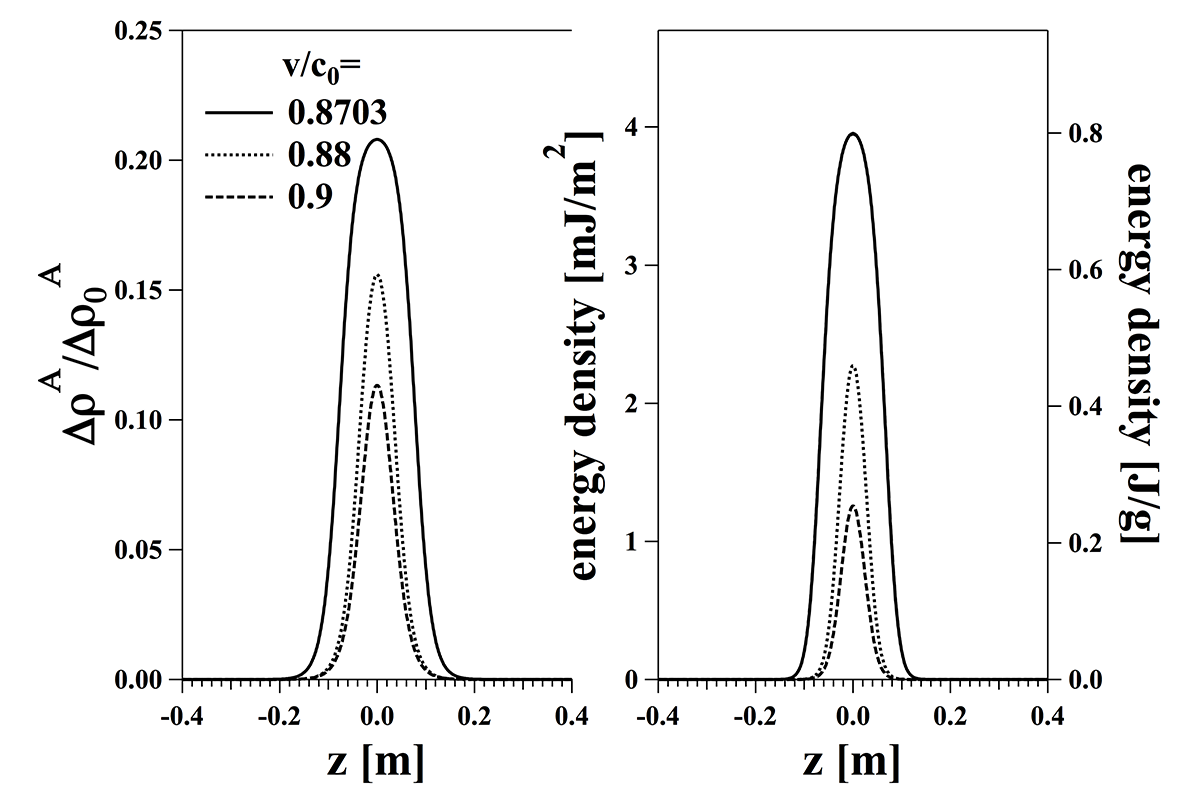}
		\parbox[c]{8cm}{ \caption{\textit{Calculated soliton shapes and the corresponding energy density calculated for lung surfactant (see \cite{Heimburg2005c}) with an assumed physiological temperature of 37$^\circ$C. The three profiles correspond to different soliton velocities. They display different shapes.}
		\label{Figure_01}}}
	\end{center}
\end{figure}

\begin{figure}[htb!]
	\begin{center}
		\includegraphics[width=6.0cm]{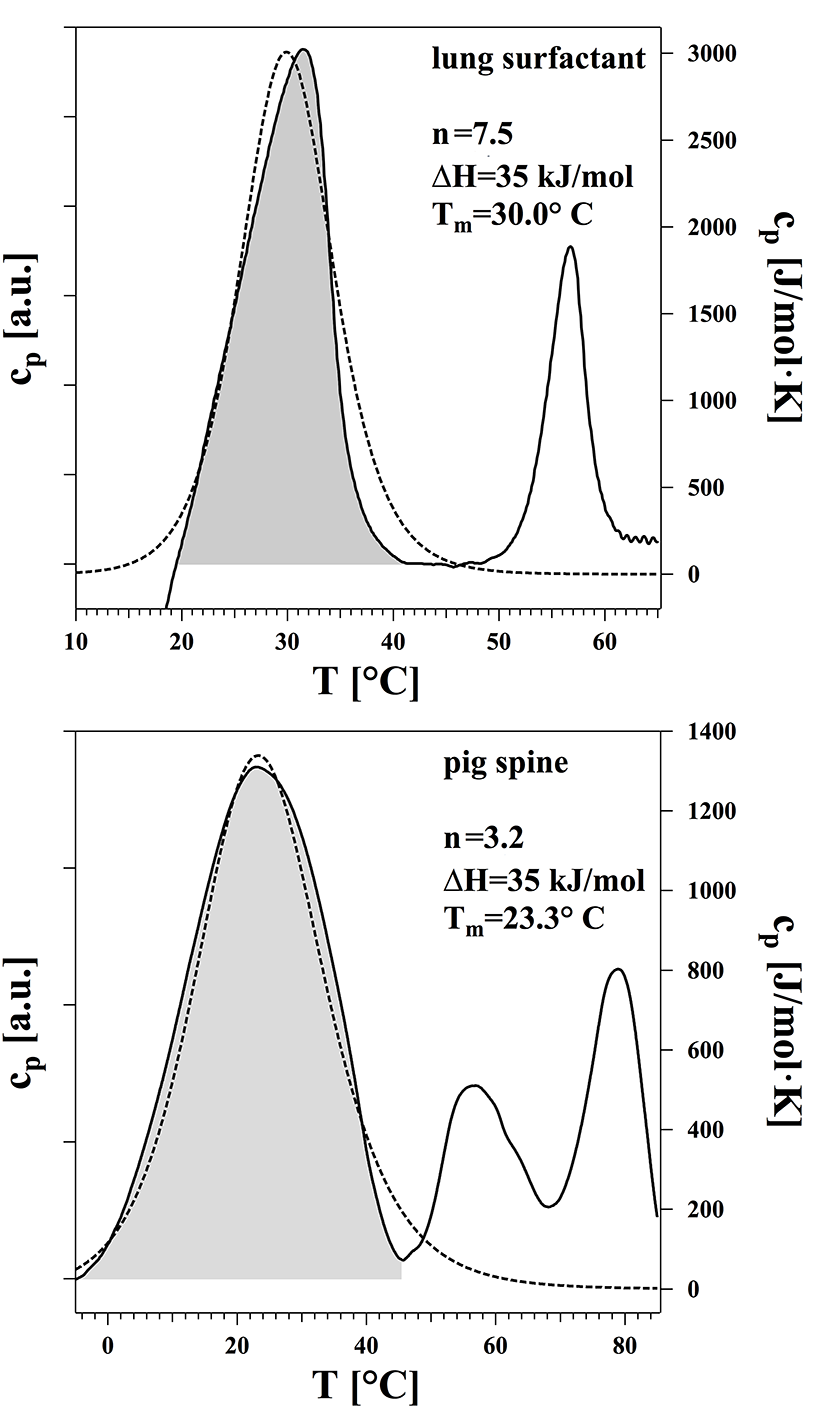}
		\parbox[c]{8cm}{ \caption{\textit{Experimental heat capacity profiles (solid lines) of lung surfactant (top, adapted from \cite{Heimburg2005c}) and native membranes from porcine spinal cord (bottom). The lipid melting peak is shaded in grey. The peaks at higher temperature represent protein unfolding transitions. The experimental profile is approximated by a van't Hoff-profile (dashed lines) assuming a transition enthalpy of 35 kJ/mol, a transition temperature of 30.0\,$^{\circ}$\,C for lung surfactant and 23.9\,$^{\circ}$\,C for porcine spinal cord. The cooperative unit sizes of are n=7.5 and n=3.2, respectively.}
		\label{Figure_02}}}
	\end{center}
\end{figure}

The hydrodynamic theory for solitary pulses in biomembranes is based on the following nonlinear differential equation \cite{Heimburg2005c}
\begin{eqnarray}\label{eq_Theory_1}
\frac{\partial^2 \Delta \rho^A}{\partial t^2}	&=&\frac{\partial  }{\partial x}\left(\left(c_0^2+p\Delta \rho^A+q(\Delta \rho^A)^2+...\right) \frac{\partial \Delta \rho^A }{\partial x}\right) \nonumber\\
&&-h\frac{\partial^4 \Delta \rho^A}{\partial x^4}\;,
\end{eqnarray}
where $\Delta \rho^A$ is the variation of the lateral density of the membrane from the equilibrium density $\rho_0^A$, and $c^2=\left(c_0^2+p\Delta \rho^A \right.$ $\left.+q(\Delta \rho^A)^2+...\right)$ describes the density dependence of the sound velocity. The parameters $c_0$, $p$ and $q$ effectively describe the density dependence of the membrane melting transition, and the dispersion parameter $h$ describes the frequency dependence of the sound velocity. The solitons have different amplitudes and energy content for different velocities $v < c_0$, where $c_0$ is the sound velocity of the membrane in the fluid phase at physiological temperature. The energy density of the soliton is given by \cite{Heimburg2005c}
\begin{equation}\label{eq_Theory_2}
e=\frac{c_0^2}{\rho_0^A}\left(\Delta \rho^A\right)^2+\frac{p}{3\rho_0^A}\left(\Delta \rho^A\right)^3+\frac{q}{6\rho_0^A}\left(\Delta \rho^A\right)^4 \;.
\end{equation}
where $\rho_0^A$ is the area density at physiological temperature. Eq. \ref{eq_Theory_1} possesses an analytical solution given in \cite{Lautrup2011}. 
The parameters in eqs. (\ref{eq_Theory_1}) and (\ref{eq_Theory_2}) must either be measured experimentally or deduced from the heat capacity. Fig.\,\ref{Figure_01} shows solitons calculated from the heat capacity profile of bovine lung surfactant in Fig.\ref{Figure_02}\,(left). The parameters for lung surfactant given in \cite{Heimburg2005c} were all calculated from the heat capacity profile and are $c_0=171.4$ m/s, $\rho_0^A=4.107 \cdot 10^{-3}$ g/m$^2$,  $p=-6.86\; \rho_0^A/c_0^A$, $q=32.32\; \rho_0^A/(c_0^A)^2$, assuming a physiological temperature of T=37$^\circ$C (in the liquid membrane phase). The dispersion parameter $h$ has no influence on the shape but only on the width of the pulse \cite{Heimburg2005c}. For this calculation it was assumed to be h=2 m$^4$/s$^2$ in order to mimic the width of an action potential. The change in density shown in Fig. \ref{Figure_01} corresponds to an increase in the chain order of the membrane. The maximum relative density change reflects the density in the solid phase.

The solitons in Fig. \ref{Figure_01} can be excited by providing free energy during the stimulation of a nerve. The aim of this paper is to calculate this free energy, and to determine its dependence on the concentration of anesthetics in the membrane. In order to do so, we wish to determine the free energy difference between a membrane at physiological temperature (which is in its liquid state) and a solid membrane at the pulse maximum.

\subsection*{Heat capacity and van't Hoff profiles}
Fig. \ref{Figure_02} shows the experimental heat capacity profile of bovine lung surfactant (taken from \cite{Heimburg2005c}) and porcine spine membranes (this work). Since it is not easy to determine the exact amount of lipid in the preparations, we use arbitrary heat capacity units. The profiles show a major peak at $T=30^{\circ}$\,C for lung surfactant, and $T=23.3^{\circ}$\,C for porcine spine, which is below the average rectal temperature of about 38.3$\,\pm$\,0.6$^{\circ}$\,C in cows, and 39.3$\,\pm$\,0.5$^{\circ}$\,C in pigs \cite{Reece2015}. These peaks correspond to lipid melting. The peaks above body temperature in the two preparations correspond to protein unfolding. They disappear in a second heating scan. Details will be reported in a separate publication focused on melting transitions in biological membrane preparations \cite{Muzic2018}. In order to simplify the calculations that follow, we approximate the lipid melting with a simulated profile. Using van't Hoff's law, the heat capacity of a lipid membrane with a cooperative unit size of $n$ can be calculated as \cite{Heimburg2007a}
\begin{eqnarray}\label{eq_Theory_3}
c_p&=&\frac{K}{(1+K)^2}\cdot \frac{n\cdot \Delta H^2}{RT^2} \nonumber\\
&\mbox{with}& \quad K=\exp\left( -n\cdot\frac{\Delta H-T\Delta S}{RT}\right) 
\end{eqnarray}
A cooperative unit size indicates that $n$ lipids undergo a transition in a cooperative manner. Fig. \ref{Figure_02} show the calculated profiles as a dashed line assuming an excess melting enthalpy of 35 kJ/mol (similar to the artificial lipid DPPC), a melting temperature of 30$^\circ$\,C for lung surfactant and 23.25$^\circ$\,C for porcine spinal cord. The cooperative unit size of $n=7.5$ and $n=3.2$, respectively. It can be seen that the transition profile for porcine spinal cord membranes is broader than that for lung surfactant. This can be due either to a different cooperative unit size, a smaller overall transition enthalpy or an imperfect purification of the membranes. For the following considerations, it is only necessary to know the heat capacity profile. We will use a profile with a transition enthalpy of 35 kJ/mol, the molar enthalpy of DPPC, a transition temperature of 296.4 K (as in porcine spine) and a cooperative unit size of $n=5.5$ (between that of lung surfactant and pocine spine).
\begin{figure*}[b!]
	\begin{center}
		\includegraphics[width=14cm]{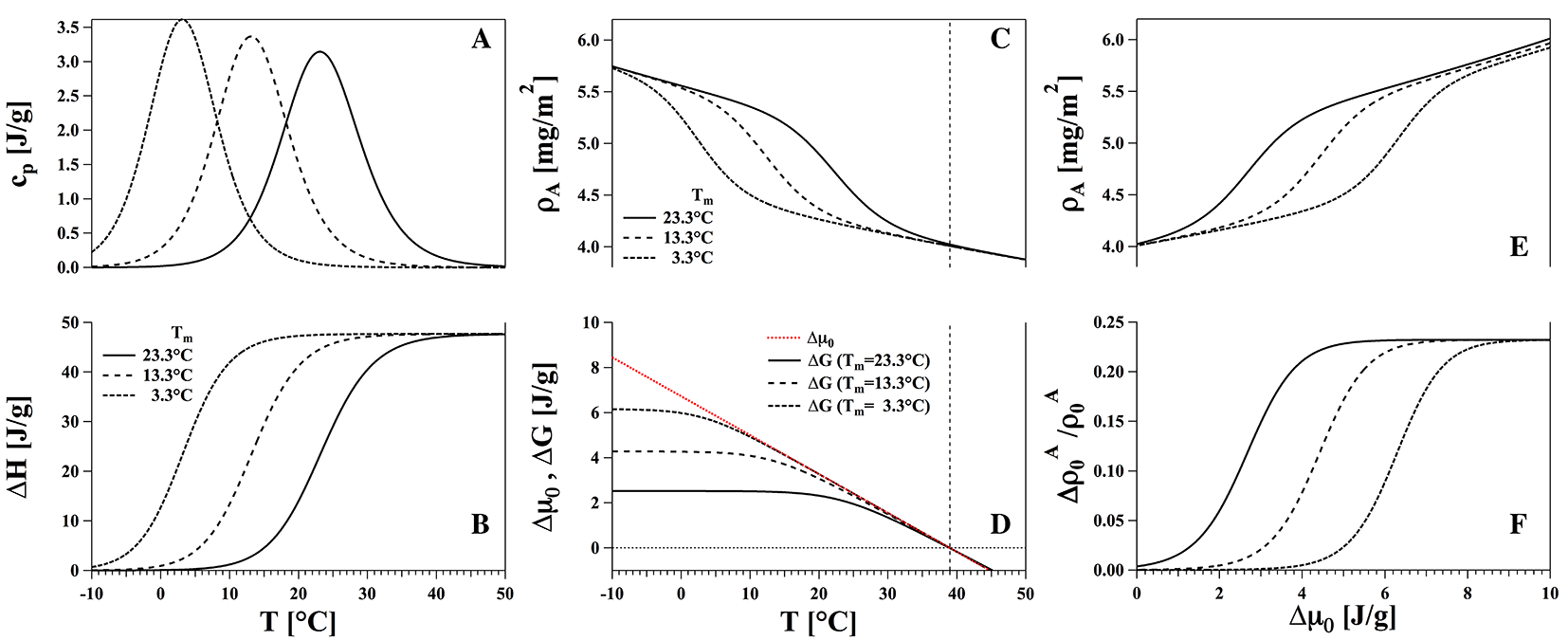}
		\parbox[c]{14cm}{ \caption{\textit{Left: Melting profiles (van't Hoff) calculated for an enthalpy of $\Delta H=$35 kJ/mol, $T_m=$23.3, 13.3 and 3.3°C, and n=5.5. Center: Area density and chemical potential differences between fluid and gel (red line) and the free energy of the membrane (black lines) assuming an experimental temperature of 39.3°C. Right: Area density and relative density change as a function of the chemical potential difference.}
		\label{Figure_03}}}
	\end{center}
\end{figure*}
This profile will be used in the following to determine the influence of anesthetics on the free energy of the membrane transition. The enthalpy and the entropy of the membrane relative to the gel phase can be determined from the heat capacity:
\begin{equation}\label{eq_Theory_4}
\Delta H(T)=\int_{T_0}^T c_p dT \quad \mbox{and} \quad \Delta S(T)=\int_{T_0}^T \frac{c_p}{T} dT \;.
\end{equation}
The chemical potential difference between liquid and solid is:
\begin{equation}\label{eq_Theory_5}
\Delta \mu_0(T)=\left(\Delta H_0-T\cdot \Delta S_0 \right)  \;,
\end{equation}
where $H_0$ and $S_0$ are the total molar transition enthalpy and entropy, respectively. The fraction of fluid lipid in the membrane is
\begin{equation}\label{eq_Theory_6}
f=\left(\frac{K}{1-K}\right) \quad\mbox{with} \quad K=\exp{\left(-\frac{n\cdot\Delta \mu_0}{kT}\right)} \;,
\end{equation}
and the free energy difference of the membrane compared to the gel state is
\begin{equation}\label{eq_Theory_7}
\Delta G(T)=f \cdot \Delta \mu_0  \;.
\end{equation}
In a solitary pulse, the area density increases and the specific area decreases. The specific area  has been calculated according to the following:
\begin{eqnarray}\label{eq_Theory_8}
A(T)&=&(1-f)\cdot A_g^{25^\circ}\cdot(1+\alpha_T \cdot (T-25^\circ)) \nonumber\\
&&+f\cdot A_f^{50^\circ}\cdot(1+\alpha_T \cdot(T-50))
\end{eqnarray}
where $f$ is the fluid lipid fraction. $A_g^{25^\circ}=194$m$^2$/g is the specific gel area of DPPC at 25$^\circ$C \cite{Sun1996},  $A_f^{50^\circ}=258.0$m$^2$/g is the specific fluid area of DPPC at 50$^\circ$C \cite{Nagle1993, Nagle1996} and $\alpha_T^A=0.003\, \mbox{K}^{-1}$ is the area expansion coefficient of DPPC, assumed to be temperature independent \cite{Kucerka2011} (values for DPPC taken from literature \cite{Heimburg1998}). In eq. (\ref{eq_Theory_8}), the temperature $T$ is given in degrees centigrade. The density $\rho^A(T) =1/A(T)$ is the inverse of the specific area.

\section*{Results}
\subsection*{Theoretical results}
Fig.\,\ref{Figure_03}A shows three heat capacity profiles for a membrane with a melting enthalpy of 35 kJ/mol and three different transition temperatures of 23.3$^\circ$C, 13.3$^\circ$C and 3.3$^\circ$C calculated for a cooperative unit size of n=5.5, which is close to that necessary to fit experimental profiles of lung surfactant and porcine spine membranes. Integration of these profiles yields $\Delta H(T)$ (Fig.\,\ref{Figure_03}B), $\Delta S(T)$ and the fluid fraction $f$ (eqs. (\ref{eq_Theory_4}) and (\ref{eq_Theory_6})). The specific area density of the membrane can be determined by using eq. (\ref{eq_Theory_8}) (Fig.\,\ref{Figure_03}C).

We can now calculate the chemical potential difference of a membrane as a function of temperature relative to the chemical potential at 39.3$^\circ$C, which is the physiological temperature in a pig (red line in Fig.\,\ref{Figure_03}D) and the free energy differences for the three profiles (black profiles in Fig.\,\ref{Figure_03}D). It can be seen that the threshold for the density change depends on the melting temperature. The absolute area density $\rho^A$ of the membrane and the relative density change ($\Delta \rho^A/\rho_0^A$) can be plotted as a function of the chemical potential difference (Fig.\,\ref{Figure_03}\,E and F). Here, $\rho_0^A$ is the density at physiological temperature.
\begin{figure*}[b!]
	\begin{center}
		\includegraphics[width=14cm]{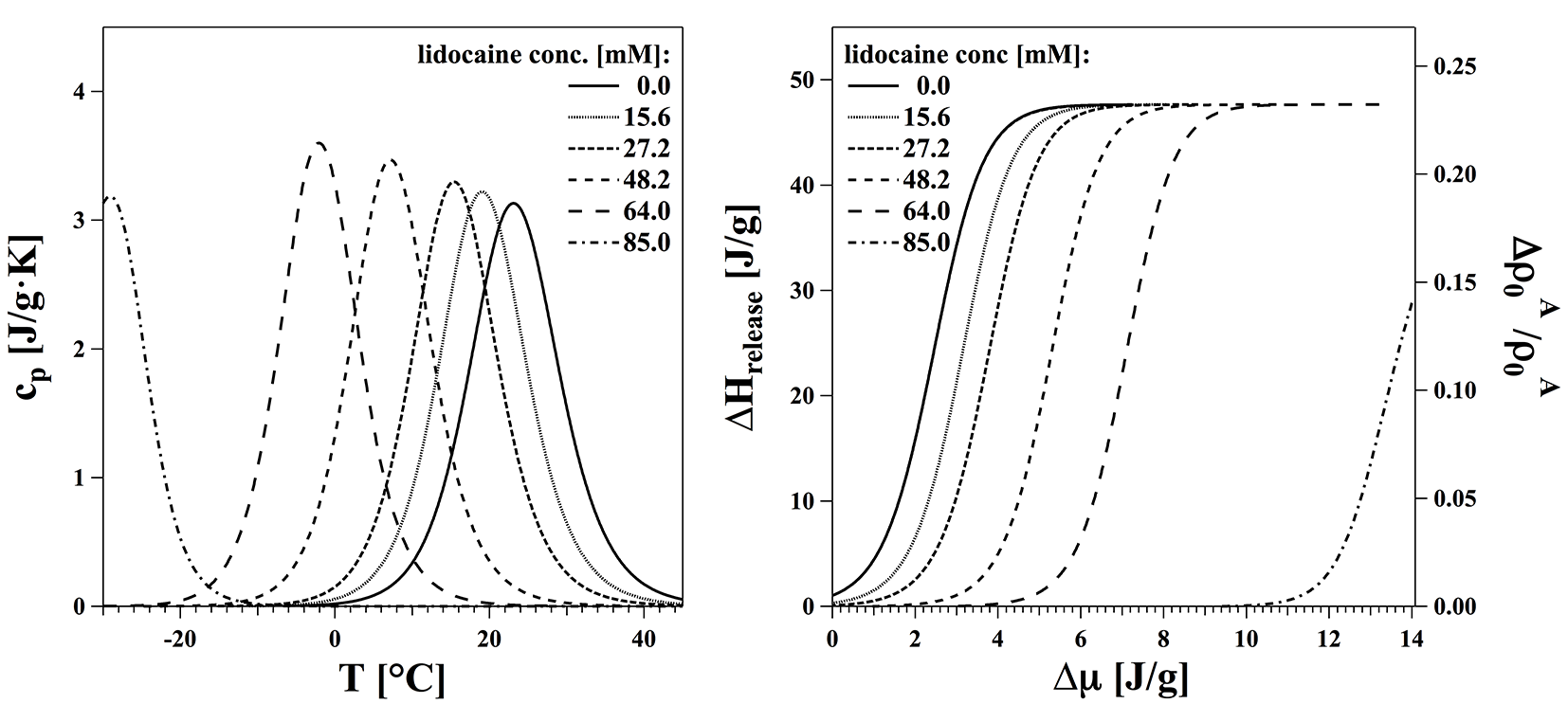}
		\parbox[c]{14cm}{ \caption{\textit{Left: Van't Hoff profiles of a membrane with $\Delta H_0=35$ kJ/mol,, $T_m=$23.3$^{\circ}$C and a cooperative unit size of n=5.5 in the presence of six different bulk concentrations of lidocaine. Right: Excess enthalpy and relative change in area density as a function of the chemical potential difference between gel and fluid membrane relative to the chemical potential difference at an experimental temperature of 39.3$^{\circ}$C (body temperature).  }
		\label{Figure_04}}}
	\end{center}
\end{figure*}
\subsection*{Anesthetics}
The transition temperature of a membrane can be influenced by the presence of anesthetics. We will treat the case of the local anesthetic lidocaine. For this anesthetic, the partition coefficient is known. (It is in the range between 10-40, depending on the nature of the membrane \cite{Graesboll2014}. Here, we use P=17.) Further, the concentration dependence of the transition temperature is known. In \cite{Graesboll2014} we have calculated the dependence of theoretical heat capacity profiles as a function of lidocaine concentration and compared these profiles with experiment. We use the same procedure here to determine the dependence of a heat capacity profile with a maximum at 23.3$^\circ$C in the absence of anesthetics (as in the porcine spinal cord) at six different lidocaine concentrations (0mM, 15.56mM, 27.20mM, 48.16mM, 63.98mM and 85.00 mM). Later in this paper, we will compare the calculated results with experiments on the median nerve of humans in which high doses of anesthetics were administered. The stimulus-response profiles were recorded over a period of several hours, during which anesthetic concentration decays as a function of time. In order to relate the above concentrations to a time after administration of anesthetics as obtained in clinical experiments by Moldovan et al. \cite{Moldovan2014} (see Fig. \ref{Figure_06}, right), we assume a time dependence of the anesthetics concentration given by
\begin{equation}\label{eq_Results_8}
c_{lido}=85 \mbox{[mM]}\cdot \exp(-2.8402\cdot\mbox{time[h]}) \;.
\end{equation}
The above concentrations now correspond to a profile without anesthetics (pre) and profiles after 0h, 1h, 2h, 4h, 6h after administration of the drug (Fig. \ref{Figure_06}). It should be noted that the melting-point depression induced by anesthetics, $\Delta T_m=-(RT_m^2/\Delta H)\cdot x_A$, depends on the melting enthalpy of the lipid membrane and on the exact value of the partition coefficient. These values are not known exactly.  In order to get an order of magnitude estimation, we have assumed that $\Delta H=35$ kJ/mol (as in DPPC) and used a partition coefficient of $P=17$ taken from \cite{Graesboll2014}. However, the true enthalpy could be significantly smaller and the partition coefficient higher. This would also render the anesthetics concentration required to induce a given shift in the melting profile significantly smaller. The values used here should thus be considered as a proof-of-principle calculation of the qualitative effects.

Figure \ref{Figure_04}\,(left) shows the heat capacity profiles for the six different lidocaine concentrations. As shown above, the excess enthalpy and entropy, and the chemical potential difference between gel and fluid membrane can be calculated by integrating the heat capacity profile. One can further determine the relative change in area density by using eq. \ref{eq_Theory_8}. Figure \ref{Figure_04}\,(right) shows the excess enthalpy and the relative density change as a function of the chemical potential difference. As a reference temperature for the chemical potential we have taken 39.3$^\circ$C, which corresponds to the body temperature of pigs. It can be seen that the curves shift towards higher potentials upon increasing the concentration of lidocaine.  We will identify an increasing chemical potential difference at threshold (50\% of full amplitude) with an increasing free energy necessary to create a soliton. The $\Delta H_{\mbox{\footnotesize release}}$ in the right hand panel corresponds to the large heat transfer from the membrane to the bulk water phase that has been observed in nerve experiments \cite{Abbott1958, Howarth1968, Howarth1975, Ritchie1985}.

We see that the threshold for the induction of density chan\-ges in the membrane depends on the chemical potential (and the Gibbs free energy). With increasing anesthetic concentration, the threshold increases (in the present example by as much as 7-fold). 

\subsection*{Experimental results}
In the preceding theory section we predicted that the thres\-hold for nerve stimulation will increase upon addition of general or local anesthetics without significantly changing the amplitude of the pulse. In the electromechanical concept, there will be no anesthetic concentration that blocks the nerve to a degree that it cannot be stimulated. In the thermodynamic approach, a pulse can always be created by further increasing the stimulus.  \\*[0.1cm]
\begin{figure*}[b!]
	\begin{center}
		\includegraphics[width=10cm]{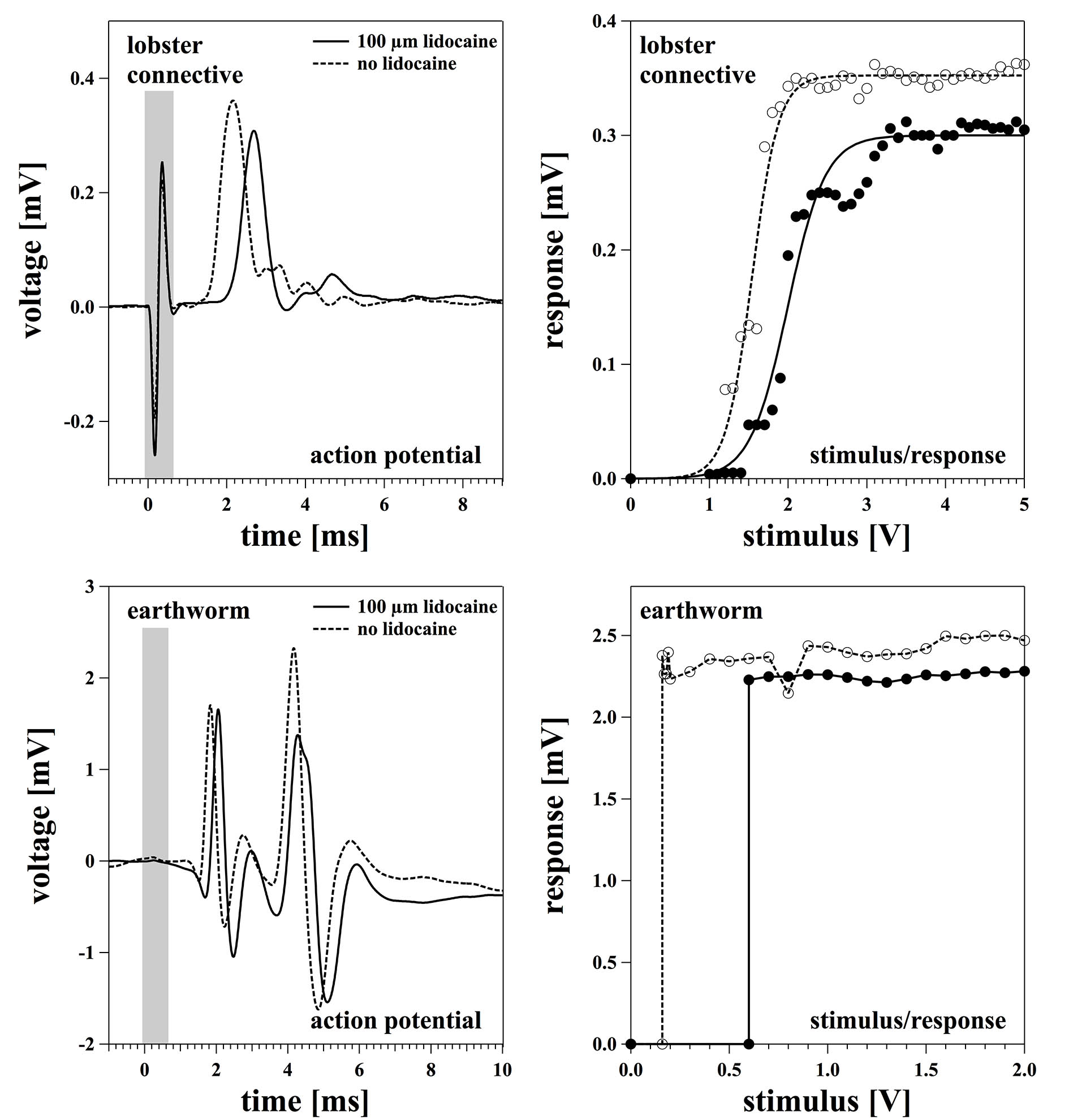}
		\parbox[c]{14cm}{ \caption{\textit{Action potentials of lobster connectives and and earthworm, and the associated stimulus-response curves in the presence and the absence of 100 \textmu M lidocaine. In both cases, the stimulus-response curve is shifted towards higher thresholds without changing the amplitude significantly.}
		\label{Figure_05}}}
	\end{center}
\end{figure*}
\noindent\textbf{Lobster cumoesophageal connectives: }Typical action potentials recorded from the connectives before (in lobster saline solution, see Materials and Methods) and after treatment with lidocaine (saline with 100\,\textmu M lidocaine) are shown in Fig. \ref{Figure_05}\,(top left). The sharp peaks in the grey-shaded regions represent the stimulation artifact. The peaks from the nerve impulse are compound action potentials (CAPs) from different groups of axons. We can directly identify two features of the first CAP after addition of lidocaine. First, there is only a slight decrease in the peak amplitude; second, the latency between stimulation and the peak increases, indicating a decrease of conduction velocity. The peak amplitude (response) is then plotted as a function of stimulation voltage and shown in Fig. \ref{Figure_05}\,(top right).  We can recognize a shift of the stimulus-response curve towards higher stimulation voltages by about 40\%. \\*[0.1cm]
\begin{figure*}[b!]
	\begin{center}
		\includegraphics[width=14cm]{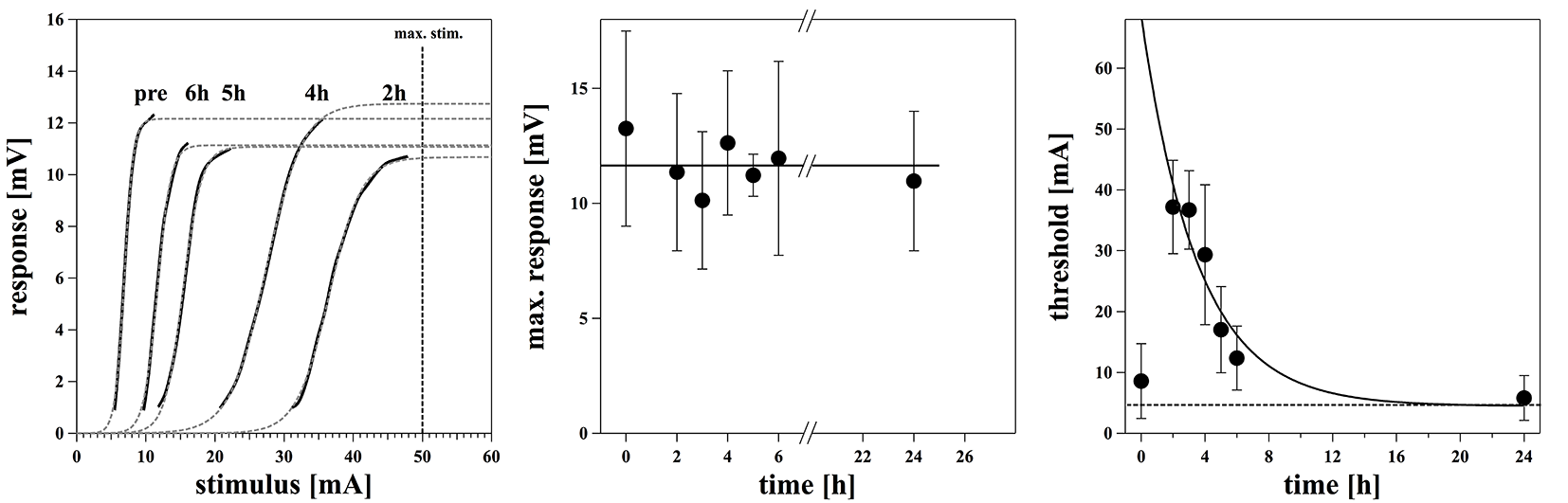}
		\parbox[c]{14cm}{ \caption{\textit{Anesthesia of the human median nerve \cite{Moldovan2014}. Left: Stimulus-response profile for the median nerve in humans ofter lidocaine anesthesia, measured at different times after application of the local anesthetic (dashed profiles represent guides to the eye). Center: The mean amplitude of the response for different times after application of lidocaine. Right: Threshold for 50\% response in the stimulus-response curves. The solid line represents an exponential fit to the data. The experimental data were adapted from Moldovan et al. \cite{Moldovan2014}.}
		\label{Figure_06}}}
	\end{center}
\end{figure*}
\noindent\textbf{Earthworm ventral cord: } The ventral cord from earthworms contains three giant neurons (one median giant axon and two lateral giant axons). The two lateral giant axons are connected to each other at each ganglion (see \cite{GonzalezPerez2014} for details). Two peaks are present in the electrophysiological recording \ref{Figure_05}\,(bottom left). The first peak corresponds to the action potential from the median giant axon, which is adopted here as representative of the response of a single neuron, and the second peak is a compound action potential from the two lateral giants. The stimulation artifact (grey-shaded area) at 0 s can barely be seen as the action potentials from earthworm ventral cord are significantly larger than those from the lobster nerves. The latency between the stimulation and the action potential increased slightly for both peaks, indicating a slight decrease in conduction velocity. In Fig.\,\ref{Figure_05}\,(bottom right), the amplitude of the first peak is plotted as a function of stimulation voltage. Since the excitation of single neurons in earthworm appears as an all-or-nothing phenomenon, the plot shows a discontinuity in the response amplitude. The threshold voltage increased by about a factor of 4. \\*[0.1cm]
\noindent\textbf{Human median nerve: } The data in Fig. \ref{Figure_06} were extracted from a publication of Moldovan et al. \cite{Moldovan2014}. Two of the authors of the present study were present during the experiments. In \cite{Moldovan2014}, the stimulus-response profiles of the human median nerve were measured before and after administration of anesthetics. The stimulation was made with two electrodes on the lower arm, and the response was measured by motor nerves in the thumb (abductor pollicis brevis, APB). The results in this study (Fig.\,\ref{Figure_06}) represent averages over several subjects.  \\*[0.1cm]
In \cite{Moldovan2014}, lidocaine was administered with syringes at several injection points directly at the median nerve. The response was followed for up to 24 hours (Fig.\,\ref{Figure_06}, left). One finds sigmoidal stimulus-response profiles with a stimulus\linebreak threshold value define by half-maximum amplitude of the response. Directly before the administration of lidocaine, the threshold stimulus current was about 8.5 mA. Directly after administration of the anesthetic, the threshold current could not be measured because it was above the maximum current that was allowed in the procedure (above 50mA). With time, a stimulation-response profile returned, probably due to passive diffusion or active transport of anesthetics away from the median nerve. Two hours after administration of the anesthetics, the threshold was 37 mA. It decayed to 29 mA after 4 hours, and to 12.4 mA after 6 hours. After 24 hours, the threshold was 5.8 mA.  This indicates that the change in threshold two hours after administration of the anesthetics was 6.4 times higher than the threshold after 24 hours. Fig.\ref{Figure_06}\,(left) should be compared to the theoretical result in Fig.\,\ref{Figure_04}. Throughout the experiment, the maximum amplitude of the response was practically unchanged within error (Fig.\ref{Figure_06}, center). We fitted the decay of the threshold with an exponential given by eq.\,(\ref{eq_Results_8}) (Fig.\ref{Figure_06}, right). This function was used to determine anesthetic concentrations in the calculations leading to Fig.\,\ref{Figure_04}.  \\*[0.1cm]

In summary, the overall behavior of the stimulus-response profiles, both in lobster, earthworm and the human median nerve is very similar to the theoretical results shown in Fig.4, that are based on the thermodynamics of biomembrane transitions.


\section*{Discussion}
During the past decade, we have demonstrated that nerve pulses have many features characteristic of the electromechanical solitons described by eq.\,(\ref{eq_Theory_1}) \cite{Heimburg2005c, Heimburg2007b, Heimburg2008, Andersen2009, Lautrup2011}.  For example, it has been shown experimentally that the action potential exchanges heat reversibly with the nerve environment \cite{Abbott1958, Howarth1968, Howarth1975, Ritchie1985, Tasaki1989}.  This indicates a reversible process such as the propagation of sound \cite{Hill1912, Heimburg2017}. In the soliton picture, such heat changes correspond to the latent heat of the membrane transition that is temporarily transferred to the membrane environment during the pulse. The list of such similarities is so large that we were led to propose that the mechanical pulse and the action potential are based on the same physics. One purpose of the present paper is to demonstrate that this assumption leads to predictions for stimulus-response profiles that resemble those obtained from living nerves. 

Under physiological conditions, the biomembrane is in its fluid state slightly above a melting transition of the membrane lipids (Fig. \ref{Figure_02}).  Under these conditions, the low frequency sound velocity of a longitudinal compressional wave along the membrane plane is of the order of 170 m/s \cite{Heimburg2005c}. However, the sound velocity depends on the magnitude of the lateral compressibility, which displays a maximum in the phase transition from ordered to disordered lipids \cite{Halstenberg1998, Heimburg1998, Heimburg2005c}, i.e., it depends on temperature and other intensive variables such as lateral pressure and transmembrane voltage. This fact gives rise to the possibility of solitary pulse propagation with a velocity that is smaller than the speed of sound. Within the  pulse, the membrane is reversibly shifted through its transition from fluid to gel (and back) with associated changes in lateral density,  polarization, and enthalpy. This gives rise to changes in the transmembrane voltage \cite{Heimburg2012, Mosgaard2015a} and to heat exchange with the membrane environment \cite{Mosgaard2013a}.

The response of a membrane to a local perturbation depends on the distance of the phase transition from ambient conditions (i.e., distance in temperature, pressure, voltage, etc.). The threshold for the induction of the phase transition is also the threshold for the generation of a solitary pulse.  Here, we monitored the stimulation threshold by observing both the lateral mass density change of the membrane, and the enthalpy change associated to it. The density change coincides with the variation in membrane thickness measured during the action potential \cite{Iwasa1980a, Iwasa1980b, Tasaki1989, GonzalezPerez2016}, while the enthalpy corresponds to the heat exchanged between membrane and environment during the pulse \cite{Abbott1958, Howarth1968, Howarth1975, Ritchie1985, Tasaki1989}. We find a sigmoidal change in the density as a function of stimulation free energy (Fig. \ref{Figure_03}). We designate the half height of this profile as the `threshold' whose value corresponds to the distance between ambient conditions and the phase transition. This threshold necessarily increases with the addition of anesthetics since the transition temperature decreases. Changes in threshold can be very large at high anesthetic concentrations, but there does not appear to be a value where the pulse is effectively blocked. The maximum amplitude of the pulse is dictated by the density difference between gel and fluid membrane and remains constant for a stimulus above threshold.

Pulse propagation has been observed experimentally in artificial lipid monolayers. This can be considered as a proof-of-principle for the above concepts. Griesbauer et al. showed that local perturbations can produced localized solitary pulses on lipid monolayers close to their `liquid expanded' to `liquid condensed' (LE-LC) transition \cite{Griesbauer2012a, Griesbauer2012b}. These pulses could be seen not only as a variation in the local pressure but could also be detected as a voltage pulse as a consequence of dipolar rearrangements of the lipid headgroups in the transition. Shrivastava et al. \cite{Shrivastava2014, Shrivastava2015} perturbed a monolayer by a local stimulus and showed that the emergence of pulses and their threshold value depends on the distance from the phase transition. The amplitude of the response could be altered by changing the lateral pressure of the film and thereby the distance between ambient conditions and the LE-LC transition. This is precisely the mechanism proposed in \cite{Heimburg2005c, Heimburg2007b, Heimburg2007c} and explored here. The response of the membrane to a perturbation in these experiments follows the theoretical profiles described in Fig. \ref{Figure_04}.

The effects of local anesthesia are often described as due to the blocking of sodium channels, but it is widely believed that the nature of general anesthesia is not understood. Interestingly,  however, general anesthesia follows the simple regularities of the well-established Meyer-Overton correlation \cite{Overton1991} which states that the potency of an anesthetic is proportional to its solubility in the lipid membrane but is otherwise completely independent of the chemistry of the drug \cite{Heimburg2007c}.  This correlation alone does not provide a particular mechanism for anesthesia.  However, if one assumes that anesthetics are perfectly miscible in the fluid membrane and perfectly immiscible in the gel phase, one obtains melting point depression, which is a general law of physical chemistry that is also independent of the chemical nature of the anesthetic.  Melting-point depression of anesthetics has been verified experimentally for substances that follow the Meyer-Overton correlation (e.g., \cite{Kharakoz2001, Heimburg2007c, Graesboll2014}). As we have shown in this paper, melting-point depression provides a natural mechanism for the action of anesthetics given the solitonic nature of the nerve pulse. We have shown in \cite{Heimburg2007c} that melting-point depression is consistent with the finding of the pressure reversal of anesthesia \cite{Johnson1950}.  We have also shown that local anesthetics have effects on lipid membranes that are similar or identical to those due to general anesthetics, e.g., they induce melting-point depression \cite{Graesboll2014}. This suggests that local anesthetics cause pronounced effects on lipid membranes without binding specifically to a target molecule. In this work we have focused on the local anesthetic lidocaine. As in \cite{Graesboll2014}, we assume that its influence on melting-points is identical to that of general anesthetics. This view is supported by the finding that local and general anesthetics display an additive effect \cite{DiFasio1976, Himes1977, Ben-Shlomo1997, Ben-Shlomo2003, Senturk2002, Agarwal2004, Pypendop2005, Altermatt2012} suggesting a common mechanism that is in agreement with both the Meyer-Overton correlation and melting-point depression.  

We have shown here that the free energy required for the generation of a solitary pulse is increased by the presence of anesthetics (Fig. \ref{Figure_04}). We have compared this to the increase in stimulation threshold for nerve excitation in experiments on living nerves.  We have demonstrated an increase in stimulation threshold by lidocaine for both earthworm axons and lobster connectives (Fig. \ref{Figure_05}). Kassahun \& al. \cite{Kassahun2010} showed that the threshold for the stimulation of bullfrog sciatic nerve increased in the presence of both the general anesthetic xenon and the local anesthetic procaine, while the maximum amplitude remained unaltered. This similarity between the effects of xenon and procaine suggests that the action of the different classes of anesthetics is comparable. Moldovan et al. anesthetized the median nerve in the human arm with administration of lidocaine \cite{Moldovan2014} (Fig. \ref{Figure_06}). Their results indicate a constant maximum response amplitude and an increase in stimulation threshold by a factor of as much as 7 to 10.  (An upper limit could not be determined do to limitations on the stimulation current required by ethical regulations.)  It is worthwhile noting that models based on the blocking of sodium channels \cite{Butterworth1990, Fozzard2011,Noble1966} predict a decreasing maximum response amplitude and cannot explain a threshold increase of the observed magnitude. However, as shown by comparing Fig.\,\ref{Figure_04} and Fig.\,\ref{Figure_06}, the findings of \cite{Moldovan2014} closely resemble those for the soliton picture. In this picture, the maximum response amplitude stays constant while the exists no highest threshold value.  We thus find it reasonable to presume that a single mechanisms underlies the action of both general and local anesthetics and, further, that this mechanism is a non-specific consequence of physical chemistry.

\section*{Conclusions}
We have shown here that a thermodynamics picture of nerve activity is consistent with the mechanical and thermal changes observed during the nerve pulse and with the action of anesthetics. Within the soliton picture, the action of anesthetics leads to an increase in stimulation threshold very similar to that observed in experiments on earthworm axons and  lobster connectives (this study) as well in the frog sciatic nerve and the human median nerve. Together with the inherent role played by mechanical changes, temperature changes and the known pressure reversal of anesthesia, the soliton model provides a consistent and self-contained picture of nerve stimulation and anesthesia.


\vspace{0.5cm}\noindent\textbf{Acknowledgments:} We thank the authors of \cite{Moldovan2014} for allowing us to be present during their stimulation experiments on the median nerves of living humans in the years 2011-2013. This work was supported by the Villum Foundation (VKR 022130).


\footnotesize{

}


\normalsize
\end{document}